\documentclass[aps, prd, reprint, letterpaper, preprintnumbers, floatfix, superscriptaddress]{revtex4-1} 

\usepackage{amsmath}
\usepackage{amssymb}
\usepackage[australian]{babel}
\usepackage{braket}
\usepackage{graphicx}
\usepackage{hyperref}
\usepackage[utf8]{inputenc}\DeclareUnicodeCharacter{2212}{-}
\usepackage{mathtools}
\usepackage{nicefrac}
\usepackage{siunitx}
\usepackage[caption=false]{subfig}
\usepackage[normalem]{ulem}
\usepackage{wasysym}
\usepackage{xcolor}

\usepackage{amsmath}
\usepackage{amssymb}
\usepackage{amsfonts}
\usepackage{amsthm}
\usepackage{bm}
\usepackage{xfrac}

\allowdisplaybreaks[1]
\newcommand{\conteqn}{\nonumber\\*}
\newcommand{\neweqn}{\\}

\newcommand{\rme}[0]{\mathrm{e}}
\newcommand{\ii}[0]{i}

\newcommand{\identity}[0]{\mathbb{I}}

\newcommand{\vect}[1]{\bm{#1}}
\newcommand{\unitvect}[1]{\smash{\widehat{\vect{#1}}}\;\!\vphantom{#1}}

\newcommand{\clovercoeff}{C_{SW}}

\newcommand{\temporalsites}{N_t}

\newcommand{\current}{\mathcal{J}}
\newcommand{\projm}{\projector{-\vect{p}}}

\newcommand{\projmpo}{\projector{\mp\vect{p}}}
\newcommand{\projmpp}{\projector{\mp\vect{p}'}}

\newcommand{\spinhalf}[0]{spin-\sfrac{1}{2}}

\newcommand{\transform}[1]{\xrightarrow{#1}}

\newcommand{\adjoint}[1]{\smash{\overline{#1}}\vphantom{#1}}

\newcommand{\transpose}{^\top}

\DeclareMathOperator{\Tr}{Tr}

\DeclareMathOperator{\sign}{sign}

\newcommand{\definedby}{\equiv}

\newcommand{\backderivLDr}[0]{{\overleftarrow{\nabla}^{{\rho}}}}

\newcommand{\deltaLDmLDn}[0]{{\delta^{{\mu}{\nu}}}}
\newcommand{\energySi}[1]{{E^{{\alpha}}(#1)}}
\newcommand{\energySj}[1]{{E^{{\beta}}(#1)}}
\newcommand{\epsilonCaCbCc}[0]{{\epsilon^{{a}{b}{c}}}}

\newcommand{\ffStarDirac}[1][Q^2]{{F^{{*}}_{{1}}(#1)}}
\newcommand{\ffStarPauli}[1][Q^2]{{F^{{*}}_{{2}}(#1)}}
\newcommand{\forwderivLDr}[0]{{\overrightarrow{\nabla}^{{\rho}}}}
\newcommand{\gammaLCk}[0]{{\gamma^{{k}}}}
\newcommand{\gammaLUfive}[0]{{\gamma^{{5}}}}
\newcommand{\gammaLUfour}[0]{{\gamma^{{4}}}}
\newcommand{\gammaLUn}[0]{{\gamma^{{\nu}}}}
\newcommand{\gevectleftSiOi}[0]{{v^{{\alpha}}_{{i}}}}
\newcommand{\gevectrightSiOj}[0]{{u^{{\alpha}}_{{j}}}}

\newcommand{\indLDm}[0]{{\mu}}

\newcommand{\indLDn}[0]{{\nu}}

\newcommand{\indSi}[0]{{\alpha}}

\newcommand{\indSj}[0]{{\beta}}

\newcommand{\indexLCk}[1]{{{#1}^{{k}}}}
\newcommand{\indexLDm}[1]{{{#1}^{{\mu}}}}
\newcommand{\indexLDn}[1]{{{#1}^{{\nu}}}}
\newcommand{\interpOi}[2]{{\chi_{{i}}(#2)}}
\newcommand{\interpPEVAOip}[2]{{\chi_{{#1}\,{i'}}(#2)}}
\newcommand{\interpPEVAOi}[2]{{\chi_{{#1}\,{i}}(#2)}}

\newcommand{\interpbarPEVAOj}[2]{{\adjoint{\chi}_{{#1}\,{j}}(#2)}}

\newcommand{\interpoptPEVASi}[2]{{\phi^{{\alpha}\,{}}_{{#1}}(#2)}}
\newcommand{\interpoptPEVASj}[2]{{\phi^{{\beta}\,{}}_{{#1}}(#2)}}
\newcommand{\interpoptbarPEVASi}[2]{{\adjoint{\phi}^{{\alpha}\,{}}_{{#1}}(#2)}}

\newcommand{\massSi}[0]{{m^{{\alpha}}}}
\newcommand{\massSj}[0]{{m^{{\beta}}}}

\newcommand{\nucleoninterpOne}[0]{{\chi_{{1}}}}
\newcommand{\nucleoninterpTwo}[0]{{\chi_{{2}}}}
\newcommand{\projectorLCk}[0]{{\Gamma^{{k}}}}
\newcommand{\projectorLUfour}[0]{{\Gamma^{{4}}}}
\newcommand{\projectorLUn}[0]{{\Gamma^{{\nu}}}}
\newcommand{\projector}[2][]{{\Gamma^{{#1}}_{\!{#2}}}}

\newcommand{\quarka}[0]{{q}}

\newcommand{\quarkbara}[0]{{\adjoint{q}}}

\newcommand{\quarkdownCb}[0]{{d^{{b}}}}

\newcommand{\quarkupCa}[0]{{u^{{a}}}}
\newcommand{\quarkupCc}[0]{{u^{{c}}}}

\newcommand{\sigmaLUmLUn}[0]{{\sigma^{{\mu}{\nu}}}}
\newcommand{\sigmaLUrLUm}[0]{{\sigma^{{\rho}{\mu}}}}
\newcommand{\spinorSi}[1][s]{{u^{{\alpha}}{(p, #1)}}}

\newcommand{\spinorpbarSj}[1][s']{{\adjoint{u}^{{\beta}}{(p', #1)}}}

\newcommand{\threecfSjSi}[6][]{{\mathcal{G}^{{3}}_{{#1}}(#2\,;#3\,,#4\,;#5\,,#6\,;\indSi{}\transform{}\indSj{})}}

\newcommand{\threecfprojSiSj}[7][]{{G^{{3}}_{{#1}}(#7\,;#2\,;#3\,,#4\,;#5\,,#6\,;\indSj{}\transform{}\indSi{})}}

\newcommand{\threecfprojSjSi}[7][]{{G^{{3}}_{{#1}}(#7\,;#2\,;#3\,,#4\,;#5\,,#6\,;\indSi{}\transform{}\indSj{})}}

\newcommand{\twocfprojPEVAOiOj}[2]{{G_{{i}{j}}(#1\,;#2)}}

\newcommand{\twocfprojPEVASi}[2]{{G(#1\,;#2\,;\indSi{})}}

\newcommand{\twocfprojPEVASj}[2]{{G(#1\,;#2\,;\indSj{})}}

\newcommand{\twocfprojPEVA}[2]{{G(#1\,;#2)}}

\newcommand{\vectorcurrentLUm}[1][]{{j^{{\mu}}_{{#1}}}}

\allowdisplaybreaks[1]

\usepackage{slashed}

\newcommand{\atobij}{\alpha^i\rightarrow\beta^j}

\newcommand{\lvac}{\langle \Omega |}

\newcommand{\ub}{\bar{u}}

\newcommand{\cA}{\mathcal{A}}

\newcommand{\cS}{\mathcal{S}}
\newcommand{\cZ}{\mathcal{Z}}

\newcommand{\be}{\begin{equation}}
\newcommand{\ee}{\end{equation}}

\newcommand{\p}{\vect{p}}
\newcommand{\pp}{\vect{p}'}
\newcommand{\q}{\vect{q}}

\newcommand{\mathdefault}[1][]{}

\begin{document}

\preprint{ADP-24-06/T1245}

\title{Odd-Parity Nucleon Electromagnetic Transitions in Lattice QCD}
%\date{13 April 2021}

\author{Finn~M.~Stokes}
\affiliation{Special Research Centre for the Subatomic Structure of
  Matter,\\Department of Physics, University of Adelaide, South
  Australia 5005, Australia}
\affiliation{J\"ulich Supercomputing Centre, Institute for Advanced Simulation,\\
  Forschungszentrum J\"ulich, J\"ulich D-52425, Germany}
\author{Benjamin~J.~Owen}\thanks{Now at the Bureau of Meteorology, Adelaide, SA, Australia.}
\affiliation{Special Research Centre for the Subatomic Structure of
  Matter,\\Department of Physics, University of Adelaide, South
  Australia 5005, Australia}
\author{Waseem~Kamleh}
\affiliation{Special Research Centre for the Subatomic Structure of
  Matter,\\Department of Physics, University of Adelaide, South
  Australia 5005, Australia}
\author{Derek~B.~Leinweber}
\affiliation{Special Research Centre for the Subatomic Structure of
  Matter,\\Department of Physics, University of Adelaide, South
  Australia 5005, Australia}

\begin{abstract}
  The parity-expanded variational analysis (PEVA) technique enables the isolation of
  opposite-parity eigenstates at finite momentum.  The approach has been used to perform the first
  lattice QCD calculations of excited-baryon form factors. In particular, these calculations show
  that the low-lying odd-parity nucleon excitations are described well by constituent quark models at
  moderate $u$ and $d$ quark masses approaching the strange quark mass.  Herein, we extend the PEVA
  technique to establish a formalism for the determination of odd-parity nucleon electromagnetic
  transition form factors in lattice QCD. The formalism is implemented in the first calculation of
  the helicity amplitudes for transitions from the ground state nucleon to the first two odd-parity
  excitations.  Through a comparison with constituent quark model calculations of these amplitudes,
  these new results give important insight into the structure of these excitations. This work is a
  critical step towards confronting experimental electroproduction amplitudes for the \(N^*(1535)\)
  and \(N^*(1650)\) resonances with \textit{ab-initio} lattice QCD calculations.
\end{abstract}

\maketitle

\section{Introduction\label{sec:introduction}}
The study of nucleon resonances is a challenging and rich field. The structure of such states can
be probed experimentally through meson electroproduction.  Such measurements of nucleon resonance
electrocouplings have existed for over a decade, but have yet to be confronted with
\textit{ab-initio} calculations.

Lattice QCD studies of resonance electrocouplings provide not only the opportunity to confront
experiment, but can also provide important insight into the nature of baryon excitations and the
manner in which their structure changes as the quark masses take values away from the physical
regime.  Through a comparison with model calculations, one can learn the veracity of the model and
gain insight into the manner in which QCD gives rise to the observed phenomena.  

This approach has already emphasised how low-lying odd-parity nucleons are described well by
constituent quark models when the light-quark masses in the simulations take moderate values
approaching the strange quark mass~\cite{Stokes:2019zdd,Chiang:2002ah,Liu:2005wg,Sharma:2013rka}.
This finding is driving new extensions of Hamiltonian effective field theory (HEFT) calculations
\cite{%
Hall:2013qba,% Original paper
Hall:2014uca,% Lambda PRL
Liu:2015ktc,% N*(1535)
Liu:2016uzk,% N*(1440)
Liu:2016wxq,% Lambda(1405)
Wu:2017qve,% Roper constraints
%Li:2019qvh,% Partial waves
%Liu:2020foc,% Kaonic Hydrogen
%Li:2021mob,% moving frame
Abell:2021awi}
where two bare-baryon basis states are incorporated in a coupled-channel analysis of $S_{11}$
pion-nucleon scattering \cite{Abell:2023qgj,Abell:2023nex}.

Computing the electrocouplings of physical resonances from lattice QCD is a challenging and
multifaceted endeavour that requires a number of problems to be solved.  First, a formalism for the
determination of odd-parity nucleon electromagnetic transition form factors in lattice QCD needs to
be established.  This is the focus of the present investigation.  Here the parity-expanded
variational analysis (PEVA) technique \cite{Stokes:2013fgw,Stokes:2018emx,Stokes:2019zdd} is
essential to the isolation of baryon eigenstates at finite momentum.

Subsequently, contributions from multi-particle scattering states need to be understood. This will
require the use of non-local momentum-projected meson-baryon operators and excellent progress is
being made for two-point functions relevant to the mass spectrum
\cite{Lang:2012db,Andersen:2017una,Silvi:2021uya,Morningstar:2021ewk,Bulava:2022vpq,BaryonScatteringBaSc:2023zvt,BaryonScatteringBaSc:2023ori,Bulava:2023uma}.
The calculation of scattering-state contributions to baryon three-point functions relevant to
matrix elements has yet to be considered.  Once the matrix elements are determined on the lattice,
one must draw on a formalism to connect the finite-volume calculations to the infinite-volume
resonances in nature.  Here scattering state contributions will play a nontrivial role.

In this light, the study presented here focuses on the matrix elements of low-lying energy
eigenstates at larger quark masses where the states dominated by single-particle operators are the
lowest-lying states in the finite-volume spectrum.  Our aim is to establish a formalism in lattice
QCD for the determination of odd-parity nucleon electromagnetic transition form factors and
illustrate the formalism in practice through first calculations of the helicity amplitudes for
transitions from the ground state nucleon to the first two odd-parity excitations.

As such, our comparison is not with experiment, but rather with constituent quark-model
calculations of these amplitudes to gain insight into the veracity of these early model
calculations \cite{Capstick:1994ne}.  We are keen to learn if the success of the quark model in
capturing the essence of the physics for both ground state and odd-parity excited-state magnetic
moments of the nucleon is realised for electromagnetic transitions between these states.  Here,
nontrivial momentum dependencies introduce new challenges for the quark model.

Thus, this investigation focuses on the problem of accessing the relevant matrix elements in lattice
QCD and the need for careful regard to state isolation and opposite-parity contaminations.
Establishing this formalism is a critical step towards confronting experimental electroproduction
amplitudes for the \(N^*(1535)\) and \(N^*(1650)\) resonances with \textit{ab-initio} lattice QCD
calculations.

It has been well-established that variational analysis techniques provide a powerful tool for
isolating individual states in lattice QCD \cite{Michael:1985ne,Luscher:1990ck}. However,
conventional variational analyses have relied upon a naive zero-momentum parity projection which
fails at finite momentum. The recent introduction and application of the parity-expanded
variational analysis (PEVA) technique~\cite{Stokes:2013fgw,Stokes:2018emx,Stokes:2019zdd} offers a
solution to this problem. By extending a conventional variational analysis, both odd and
even-parity baryon eigenstates can be simultaneously isolated at finite momentum. In this paper, we
extend this technique to the calculation of transition form factors of baryons.

Section \ref{sec:cvf} commences with a review of the covariant vertex functions for both positive
and negative parity transition matrix elements and outlines the relationship between the associated
form factors and the helicity amplitudes of experimental interest.  Here we introduce a new
formalism and discuss its relative merits.  Section \ref{sec:peva} briefly reviews the PEVA
technique and outlines its application to odd-parity electromagnetic transitions between the
nucleon ground state and odd-parity excitations of the nucleon.  Finally, the formalism is applied
to calculate the odd-parity nucleon resonance electrocouplings and the results are compared with
legacy quark-model calculations in Sec.~\ref{sec:results}.  Section \ref{sec:summary}
concludes our presentation with a summary.

\section{Covariant Vertex Functions and Helicity Amplitudes\label{sec:cvf}}

In our previous work~\cite{Stokes:2019zdd}, we made use of variational techniques to evaluate the
elastic electromagnetic form factors for the lightest negative-parity nucleon states. While such
quantities can provide important insight into the underlying structure of these states and inform
model calculations, defining and measuring such quantities experimentally present significant
challenges. Experimentally the quantities of interest are the transition elements for the
electroproduction processes $\gamma^{*} N \rightarrow N^{*}$, specifically the transverse and
longitudinal helicity amplitudes ${\cal A}_{\frac{1}{2}}$ and ${\cal S}_{\frac{1}{2}}$.

Previous lattice calculations have focused primarily on the $\Delta \rightarrow N \gamma$
transition. The framework for such a calculation was established in Ref.~\cite{Leinweber:1992pv}
and subsequently examined comprehensively both in quenched
\cite{Alexandrou:2003ea,Alexandrou:2004xn} and full QCD
\cite{Alexandrou:2007dt,Alexandrou:2010uk}. For all of these studies, due to the choice of lattice
ensemble parameters, the $\Delta$ baryon remains the lightest state in this channel and so this
process can be examined using standard techniques. The results of \cite{Alexandrou:2010uk} show
good qualitative agreement with the experimental data, especially in the value extracted for
$E2/M1$ ratio, however the authors outline that discrepancies observed in the exact behaviour of
these amplitudes, particularly the $M1$ amplitude, highlight important chiral dynamics to this
process.

The only other nucleon transition considered to date has been the $N^*(1440) \rightarrow N \gamma$
transition \cite{Lin:2008qv,Lin:2011da}. However, as yet it is unlikely any group has properly
isolated the $N^*(1440)$ on the lattice. The current understanding is that the $N^*(1440)$ Roper
resonance is dynamically generated through rescattering in the $\pi N$, $\pi \Delta$, and $\sigma
N$ channels \cite{Liu:2016uzk,Wu:2017qve,Leinweber:2024psf}, and significant advances on accessing
multiparticle states in lattice QCD are being made \cite{Morningstar:2021ewk,Bulava:2022vpq}.
Experimental determinations indicate that the sign of the $\cA_{\frac{1}{2}}$ transition form
factor is negative at low $Q^2$. This contrasts constituent quark models which predict a positive
sign, lending further evidence the Roper resonance is not associated with a quark-model state.  The
need for at least three coupled meson-baryon channels to generate this resonance \cite{Wu:2017qve}
presents a formidable challenge to first-principles lattice QCD calculations.  However, evaluation
of this amplitude will assist in understanding the underlying dynamics of the Roper resonance and
the make up of this state, both on the lattice and in phenomenology.

A notably absent calculation is the evaluation of the odd-parity $N^*(1535) \rightarrow N \gamma$
transition. The focus of this work is the extraction of the transition form factors for such
states. We begin with the presentation of a general framework through which one can determine $N^*
\rightarrow N\gamma$ transition amplitudes for all choices and combinations of parity, all from the
same baryon correlator. Then in the following section, we utilise this framework to examine the
${\cal A}_{\frac{1}{2}}$ and ${\cal S}_{\frac{1}{2}}$ form factors of the first two negative parity
nucleon eigenstates.

Experimentally the quantities of interest are the transverse and longitudinal helicity amplitudes
${\cal A}_{\frac{1}{2}}$ and ${\cal S}_{\frac{1}{2}}$
\begin{align}
	{\cal A}_{\frac{1}{2}} &= \sqrt{\frac{2\pi\alpha}{K}} \frac{1}{e} \langle \, N^*, s'_z=\sfrac{1}{2}^+ \, | \, \epsilon^{(+)}_{\mu} J^{\mu} \, | \, N, s_z=\sfrac{1}{2}^- \, \rangle \, , \\
	{\cal S}_{\frac{1}{2}} &= \sqrt{\frac{2\pi\alpha}{K}} \frac{1}{e} \frac{|\q|}{Q} \langle \, N^*, s_z'=\sfrac{1}{2}^+ \, | \, \epsilon^{(0)}_{\mu} J^{\mu} \, | \, N, s_z=\sfrac{1}{2}^+ \, \rangle \, ,
\end{align}
where $e$ is the magnitude of the electron's charge, $\alpha$ the electromagnetic fine structure
constant and $Q = \sqrt{Q^2}$. Here $\epsilon_{\mu}$ represents the polarisation of the incoming
virtual photon, with $|\q|$ the photon's 3-momentum in the $N^*$ rest frame and 
\begin{equation}
K = \frac{M^2 - m^2}{2M}\, .
\end{equation} 
We identify $M$ with the resonance $N^*$ and $m$ with the nucleon $N$. These
amplitudes can in turn be related to matrix elements with the familiar form
\begin{align}
	&\langle N^*, p', s' | \, J^{\mu} \, | N, p, s \rangle \, = & \nonumber \\
		& \hspace*{7pt} 
		e \, \left( \frac{M \, m}{E'(\pp) \, E(\p)} \right)^{\sfrac{1}{2}}
		\ub_{(M)}(p', s') \, \Gamma(p', p) \, u_{(m)}(p,s) \, ,
\end{align}
which parametrises the interaction in terms of Lorentz covariant structures.  We note that all
spinors in this and subsequent expressions are regular Dirac spinors. Here the subscript,
$u_{(\xi)}$, labels the mass of the state for which the spinor describes. A convenient choice of
parametrisation is that first presented in Ref.~\cite{Devenish:1975jd} and nicely summarised in
Refs.~\cite{Aznauryan:2008us,Aznauryan:2011qj}, which allows us to express the normal-parity
($\sfrac{1}{2}^+ \rightarrow \sfrac{1}{2}^+$) transition as
%\begin{widetext}
\begin{align}
	\label{eq:normtransG1G2}
	&\langle N^*, p', s' | J^{\mu} | N, p, s \rangle = \nonumber \\
	&\hspace*{10pt} e \, \left( \frac{M \, m}{E'(\pp) \, E(\p)} \right)^{\sfrac{1}{2}} \, \ub_{(M)}(p', s') \, \tilde{J}^{\mu} \, u_{(m)}(p, s) \, ,
\end{align}
and abnormal-parity ($\sfrac{1}{2}^+ \rightarrow \sfrac{1}{2}^-$) transition
\begin{align}
	\label{eq:abnormtransG1G2}
	&\langle N^*, p', s' | J^{\mu} | N, p, s \rangle = \nonumber \\
	&\hspace*{10pt} e \, \left( \frac{M \, m}{E'(\pp) \, E(\p)} \right)^{\sfrac{1}{2}} \, \ub_{(M)}(p', s') \, \tilde{J}^{\mu} \, \gamma_5 \, u_{(m)}(p, s) \, ,
\end{align}
with
\begin{align}
	\tilde{J}^{\mu} = \,&- \left[ q^2 \, \gamma^{\mu} - \slashed{q} \, q^{\mu} \right] \, G_1(Q^2) \nonumber\\
	&- \left[ \left( P \cdot q \right) \, \gamma^{\mu} - \slashed{q} \, P^{\mu} \right] \, G_2(Q^2) \, ,
\end{align}
and $P = \frac{1}{2} \left( p' + p \right)$.

Using this decomposition, the helicity amplitudes can be expressed in terms of $G_1$ and $G_2$. For
normal transitions
\begin{align}
	&\cA_{\frac{1}{2}}^+(Q^2) = \nonumber \\
	&\hspace*{10pt} b_{+} \, \left[ 2Q^2 \, G_1(Q^2) - (M^2-m^2) \, G_2(Q^2) \right] \, , \\
	&\cS_{\frac{1}{2}}^+(Q^2) = \nonumber \\
	&\hspace*{10pt} b_{+} \, \frac{|\q\,|}{\sqrt{2}} \, \left[ 2 \, (M+m) \, G_1(Q^2) + (M-m) \, G_2(Q^2) \right] \, ,
\end{align}
and for abnormal transitions
\begin{align}
	&\cA_{\frac{1}{2}}^-(Q^2) = \nonumber \\
	&\hspace*{10pt} b_{-} \, \left[ 2Q^2 \, G_1(Q^2) - (M^2-m^2) \, G_2(Q^2) \right] \, , \\
	&\cS_{\frac{1}{2}}^-(Q^2) = \nonumber \\
	&b_{-} \, \frac{-|\q\,|}{\sqrt{2}} \left[ 2 \, (M-m) \, G_1(Q^2) + (M+m) \, G_2(Q^2) \right] \, ,
\end{align}
%\end{widetext}
with
\begin{equation}
	b_{\pm} = \sqrt{\frac{Q^2+(M \mp m)^2}{8\,m\,(M^2-m^2)}} \, ,
%% also b_{\pm} = e \sqrt{\frac{E \mp m}{8\,m\,K}} \, .
\label{eq:bpm}
\end{equation}

Another popular choice of parametrisation is to express the vertex in terms of Pauli-Dirac-like form factors $F_1^*(Q^2)$ and $F_2^*(Q^2)$. Taking
\begin{align}
	F_1^*(Q^2) &= Q^2 \, G_1(Q^2) \, ,\label{eqn:DiracFormFactor}\\
	F_2^*(Q^2) &= -\frac{(M^2-m^2)}{2} \, G_2(Q^2) \, ,\label{eqn:PauliFormFactor}
\end{align}
one can re-express the transition matrix elements \cite{Aznauryan:2011qj} as
\begin{widetext}
\begin{equation}
	\langle N^*_i, p', s' | J^{\mu} | N, p, s \rangle \, = e \, \left( \frac{M \, m}{E'(\pp) \, E(\p)} \right)^{\sfrac{1}{2}} \, \ub_{(M)}(p', s') \, \Gamma_i^{\mu}(p', p) \, u_{(m)}(p,s) \, ,
\end{equation}
where $i$ denotes normal, $n$, or abnormal, $a$, transitions.
For normal transitions 
\be
	\label{eq:normtransF1F2}
	\Gamma^{\mu}_{\mathrm{n}}(p',p) = \left( \delta^{\mu}_{\nu}  - \frac{q_{\nu} \, q^{\mu}}{q^2} \right) \gamma^{\nu} \, F_1^*(Q^2) + \frac{i \sigma^{\mu\nu}\,q_{\nu}}{M+m} \, F_2^*(Q^2) \, ,
\ee
and for abnormal transitions
\be
	\label{eq:abnormtransF1F2}
	\Gamma^{\mu}_{\mathrm{a}}(p',p) = \left( \delta^{\mu}_{\nu} - \frac{q_{\nu} \, q^{\mu}}{q^2} \right) \, \gamma^{\nu} \, \gamma_5 \, F_1^*(Q^2) + \frac{i \sigma^{\mu\nu}\,q_{\nu}}{M-m} \, \gamma_5 \, F_2^*(Q^2) \, .
\ee
\end{widetext}
We note that our choice of normalisation for the $F_2^*$ form factor in
the abnormal transition vertex differs from the common choice \cite{Ramalho:2011ae,Ramalho:2011fa}
by a factor of $\left( \frac{M-m}{M+m} \right)$. However, the absence of parity doubling in the
low-lying energy eigenstates of QCD admit this formalism. The advantage of our normalisation is
apparent if we consider the expressions for the helicity amplitudes in terms of the Dirac- and Pauli-like
form factors, namely
\begin{align}
	&\cA^{\pm}_{\frac{1}{2}}(Q^2) = 2 \, b_{\pm} \, \left[ F_1^*(Q^2) + F_2^*(Q^2) \right] \, , \\
	&\cS^{\pm}_{\frac{1}{2}}(Q^2) = \nonumber\\
	&\hspace*{10pt} \pm \frac{\sqrt{2} \, b_{\pm} \, (M \pm m) \, |\q\,|}{Q^2} \, \left[ F_1^*(Q^2) - \tau_{\pm} \, F_2^*(Q^2) \right] \, ,
\end{align}
where $b_{\pm}$ is defined in Eq.~(\ref{eq:bpm}) and
\begin{equation}
	\tau_{\pm} = \frac{Q^2}{(M \pm m)^2} \, ,
\end{equation}
with the $\pm$ label indicating the parity of the resonant nucleon state. 
Hence, the helicity amplitudes are proportional to generalisations of the
Sachs electric and magnetic form factors
\begin{align}
	G_E^*(Q^2) &= F_1^*(Q^2) - \frac{Q^2}{(M \pm m)^2} \, F_2^*(Q^2) \, , \\
	G_M^*(Q^2) &= F_1^*(Q^2) + F_2^*(Q^2) \, .
\end{align}

Compare this to the choice of Refs.~\cite{Ramalho:2011ae,Ramalho:2011fa,Ramalho:2023hqd} where
\begin{equation}
F_2(Q^2) = \frac{M+m}{M-m} \, F_2^*(Q^2) \, ,
\end{equation}
for the abnormal transition such that
%\begin{widetext}
\begin{align}
	& \cA^{-}_{\frac{1}{2}}(Q^2) = 2 \, b_{-} \, \left[ F_1^*(Q^2) + \frac{M-m}{M+m} \, F_2(Q^2) \right] \, , \\
	& \cS^{-}_{\frac{1}{2}}(Q^2) = \nonumber\\
	& -\frac{\sqrt{2} \, b_{-} \, (M + m) \, |\q\,|}{Q^2} \, \left[ \frac{M-m}{M+m} \, F_1^*(Q^2) - \tau_{+} \, F_2(Q^2) \right] \, .
\end{align}
In this case, the above identification with generalisations of the Sachs form factors
is not possible due to a relative factor of $\left(\frac{M+m}{M-m} \right)$ between
$F_2^*$ and $F_2$. In our analysis we shall use the decompositions given by
Eqs.~\eqref{eq:normtransF1F2} and \eqref{eq:abnormtransF1F2}.

\section{Parity Expanded Variational Analysis\label{sec:peva}}

\subsection{Two-point functions}

The process of extracting transition from factors of baryonic excited states via the PEVA technique
follows similarly to the elastic case presented in Ref~\cite{Stokes:2018emx}. We provide here a
brief summary of this process and detail the generalisations required to handle transition matrix
elements.

In this section we focus on a lattice QCD implementation of the formalism. We adopt the common
Euclidean metric \(\deltaLDmLDn{}\) and use the Hermitian Pauli representation for the Dirac
matrices. With a Euclidean metric, there is no need to distinguish between contravariant and
covariant indices.

We begin with a basis of \(n\) conventional spin-\nicefrac{1}{2} operators
\(\left\{\interpOi{}{x}\right\}\) that couple to the states of interest.
We introduce the PEVA projector~\cite{Stokes:2013fgw}
\begin{equation}
\projmpo \definedby \frac{1}{4} \left(\identity + {\gammaLUfour}\right)
\left(\identity \mp \ii {\gammaLUfive} {\gammaLCk} \indexLCk{\unitvect{p}}\right)\, ,
\end{equation}
where there is an arbitrary sign choice in whether the unit vector \(\unitvect{p}\) is chosen to be
parallel (\(+\)) or antiparallel (\(-\)) to the momentum direction.  This choice changes the sign
of the cross-parity terms, but otherwise has no effect on the analysis of the two-point
functions. However, as discussed later, it plays an important role in optimising the extraction of
transition form factors for general kinematics. With this projector, a set of basis operators is
constructed
\begin{subequations}
\begin{align}
    \interpPEVAOi{\mp\vect{p}}{x} &\definedby \projmpo \, \interpOi{}{x}\,,\neweqn
    \interpPEVAOip{\mp\vect{p}}{x} &\definedby \mp \projmpo \, \gammaLUfive{} \, \interpOi{}{x}\,.
\end{align}
\end{subequations}

We then seek an optimised set of operators \(\interpoptPEVASi{\mp\vect{p}}{x}\)
that each couple strongly to a single energy eigenstate \(\indSi\). These
optimised operators are constructed as linear combinations of the basis
operators. The optimum linear combinations are found by solving a generalised
eigenvalue problem (GEVP) with
\(\twocfprojPEVA{\vect{p}}{t+\Delta{}t}\) and \(\twocfprojPEVA{\vect{p}}{t}\),
where the correlation matrix
\begin{align}
    &\twocfprojPEVAOiOj{\vect{p}}{t} \conteqn
    &\qquad\definedby
      \Tr\left(\sum_{\vect{x}} \rme^{-\ii \vect{p}\cdot\vect{x}}
      \braket{\Omega|\,\interpPEVAOi{\mp\vect{p}}{x}\,
        \interpbarPEVAOj{\mp\vect{p}}{0}\,|\Omega} \right)\,,
\end{align}
with \(i\) and \(j\) ranging over both the primed and unprimed operators. This
process is described in detail in Ref.~\cite{Stokes:2013fgw}.

Using the optimised operators, we can construct the eigenstate-projected
two-point correlation function
\begin{align}
    &\twocfprojPEVASi{\vect{p}}{t} \conteqn
    &\qquad\definedby
      \Tr\left(\sum_{\vect{x}} \rme^{-\ii \vect{p}\cdot\vect{x}}
      \braket{\Omega|\,\interpoptPEVASi{\mp\vect{p}}{x}\,
        \interpoptbarPEVASi{\mp\vect{p}}{0}\,|\Omega} \right) \conteqn
    &\qquad=\gevectleftSiOi(\vect{p})\,
      \twocfprojPEVAOiOj{\vect{p}}{t}\, \gevectrightSiOj(\vect{p})\,.
\end{align}

\subsection{Three-point functions\label{sec:threepoint}}

The same approach is applied to lattice three-point correlation functions drawing on the
eigenvectors obtained in solving the two-point GEVP
\begin{align}
    &\threecfSjSi[\mp]{\current\!}{\vect{p}'\!}{\vect{p}}{t_2}{t_1}\conteqn
    &\qquad\definedby \sum_{\vect{x}_1,\vect{x}_2} \rme^{-\ii \vect{p}'\cdot\vect{x}_2} \,
        \rme^{\ii (\vect{p}' - \vect{p})\cdot\vect{x}_1}\conteqn
    &\qquad\qquad\quad\times\braket{\Omega|\,\interpoptPEVASj{\mp\vect{p}'}{x_2}\,
        \current(x_1)\,\interpoptbarPEVASi{-\vect{p}}{0}\,|\Omega}\,,
\end{align}
where \(\current(x)\) is some current operator\index{current operator}, which is
inserted with a momentum transfer \(\vect{q} = \vect{p}' - \vect{p}\). The
consideration of \(\threecfSjSi[+]{\current\!}{\vect{p}'\!}{\vect{p}}{t_2}{t_1}\)
(where the sink operator uses the opposite PEVA projector sign convention to
the source operator) is required to optimise the extraction of the form factors
for general kinematics. We note that it is sufficient to consider this change
of projector for the sink operator alone, leaving the source operator as
\(\interpoptbarPEVASi{-\vect{p}}{0}\) in all cases considered.

In this paper, we investigate electromagnetic transitions of the proton and
neutron by choosing the current operator \(\current(x)\) to be the vector
current\index{vector current}. In particular, we use the \(O(a)\)-improved~\cite{Martinelli:1990ny}
conserved vector current\index{vector current}
used in Ref.~\cite{Boinepalli:2006xd},
\begin{equation}
    \vectorcurrentLUm[CI]{}(x) \definedby \vectorcurrentLUm[C]{}(x)
    + \frac{r}{2}\,a\, \quarkbara{}(x)\left(\backderivLDr{} + \forwderivLDr{}\right)
            \sigmaLUrLUm{}\,\quarka{}(x)\,,
\end{equation}
where \(r\) is the Wilson parameter, and \(\vectorcurrentLUm[C]{}(x)\) is the
standard conserved vector current for the Wilson action, symmetrised about a
lattice site.
% \begin{subequations}
% \begin{align}
%     \forwderivLDm{} \quarka{}(x) \definedby
%         \frac{1}{2}\big(&\linkvarLDm{}(x)\,\quarka{}(x+\basisLDm{})\conteqn
%             &- \linkvarDagLDm{}(x-\basisLDm{})\,\quarka{}(x-\basisLDm{})\big)\neweqn
%     \quarkbara{}(x) \backderivLDm{} \definedby
%         \frac{1}{2}\big(&\quarkbara{}(x+\basisLDm{})\,\linkvarDagLDm{}(x)\conteqn
%             &- \quarkbara{}(x-\basisLDm{})\,\linkvarLDm{}(x-\basisLDm{})\big)\,.
% \end{align}
% \end{subequations}
% This current is derived from the standard
% conserved vector current for the Wilson action
% \begin{align}
%     \vectorcurrentLUm[C]{}(x) \definedby \frac{1}{4}\Big[\,
%         &\quarkbara{}(x) \,(\gammaLUm{} - r)\,
%             \linkvarLUm{}(x)\,\quarka{}(x+\basisLUm{})\conteqn
%         + \,&\quarkbara{}(x+\basisLUm{}) \,(\gammaLUm{} + r)\,
%             \linkvarDagLUm{}(x)\,\quarka{}(x)\conteqn
%         + \,&\quarkbara{}(x-\basisLUm{}) \,(\gammaLUm{} - r{})\,
%             \linkvarLUm{}(x-\basisLUm{})\,\quarka{}(x)\conteqn
%         + \,&\quarkbara{}(x) \,(\gammaLUm{} + r)\,
%             \linkvarDagLUm{}(x-\basisLUm{})\,\quarka{}(x-\basisLUm{})\Big]\,.
% \end{align}

As discussed in Section~\ref{sec:cvf}, this choice of current operator gives the
parity-conserving transition matrix element
\begin{widetext}
\begin{align}
    &\braket{\indSj^{\pm}\,; p'\,; s' | \,\vectorcurrentLUm[CI]{}\, | \indSi^{\pm}\,; p\,; s}
	= \sqrt{\frac{\massSi}{\energySi{\vect{p} }}} \,
          \sqrt{\frac{\massSj}{\energySj{\vect{p}'}}} \,
	  \spinorpbarSj{} \, \Gamma_n^{\mu}(p', p) \, \spinorSi{}\conteqn
    &\qquad = \sqrt{\frac{\massSi}{\energySi{\vect{p} }}} \,
              \sqrt{\frac{\massSj}{\energySj{\vect{p}'}}} \;
              \spinorpbarSj{} %\conteqn
    %&\qquad\qquad\quad\times 
    \left(\left(\deltaLDmLDn{} - \frac{\indexLDm{q}\indexLDn{q}}{q^2}\right) \gammaLUn{}\, \ffStarDirac
        - \frac{\sigmaLUmLUn{}\,\indexLDn{q}}{\massSj + \massSi{}} \, \ffStarPauli\right)%\conteqn
    %&\qquad\qquad\quad\times
    \spinorSi{}\,,\label{eqn:formfactors:matrixelement}
\end{align}
where \(\indSi\) and \(\indSj\) have the
same parity, and the parity-changing transition matrix element
\begin{align}
    &\braket{\indSj^{\mp}\,; p'\,; s' | \,\vectorcurrentLUm[CI]{}\, | \indSi^{\pm}\,; p\,; s}
	= \sqrt{\frac{\massSi}{\energySi{\vect{p} }}} \,
          \sqrt{\frac{\massSj}{\energySj{\vect{p}'}}} \,
	  \spinorpbarSj{} \, \Gamma_a^{\mu}(p', p) \, \spinorSi{}\conteqn
    &\qquad = \sqrt{\frac{\massSi}{\energySi{\vect{p} }}} \,
              \sqrt{\frac{\massSj}{\energySj{\vect{p}'}}} \;
              \spinorpbarSj{} %\conteqn
    %&\qquad\qquad\quad\times 
    \left(\left(\deltaLDmLDn{} - \frac{\indexLDm{q}\indexLDn{q}}{q^2}\right) \gammaLUn{}\gammaLUfive{}\, \ffStarDirac
        - \frac{\sigmaLUmLUn{}\,\indexLDn{q}}{\massSj{} - \massSi{}} \, \gammaLUfive\, \ffStarPauli\right)%\conteqn
    %&\qquad\qquad\quad\times
    \spinorSi{}\,,
\label{eq:pchanging}
\end{align}
where \(\indSi\) and \(\indSj\) have opposite parity. Note that these expressions differ from those in
Section~\ref{sec:cvf} due to conversion to the Pauli representation used in the current section. Here
\(Q^2 = \vect{q}^2 - {\left(\energySi{\vect{p}'} - \energySi{\vect{p}}\right)}^2\) is the squared four-momentum with the
conventional sign, and the invariant scalar functions \(\ffStarDirac\) and \(\ffStarPauli\) are
respectively the Dirac-\index{form factors!Dirac} and Pauli-like\index{form factors!Pauli}
transition form factors defined in Equations \ref{eqn:DiracFormFactor} and
\ref{eqn:PauliFormFactor}.

To extract our desired signal from this spinor structure, we can take the spinor trace
with some spin-structure projector\index{spin-structure projector}
\(\projector{S}\). This trace is then called the
spinor-projected three-point correlation function\index{correlation function!three point}
\begin{align}
    &\threecfprojSjSi[\mp]{\vectorcurrentLUm[CI]{}}{\vect{p}'\!}{\vect{p}}{t_2}{t_1}{\projector{S}}%\conteqn
        %&\qquad
	\definedby \Tr\!\left[\,\projector{S}\,
	\threecfSjSi[\mp]{\vectorcurrentLUm[CI]{}}{\vect{p}'\!}{\vect{p}}{t_2}{t_1}\, \right]\conteqn
	&\qquad= e^{-\energySj{\vect{p}'} (t_2 - t_1)} \, e^{-\energySi{\vect{p}} t_1} \,
	\frac{\bar{\cZ}^{\alpha}(\vect{p})}{2\energySi{\vect{p} }} \, \frac{\cZ^{\beta}(\vect{p}')}{2\energySj{\vect{p}'}} \,
	\Tr\!\left[\,\projector{S}\,\Gamma_{\mp\vect{p}'} \left( \slashed{p}+m^{\alpha} \right) \, \Gamma^{\mu}_i(p,p') \, \left( \slashed{p}'+m^{\beta} \right)\,\Gamma_{-\vect{p}}\,\right] \,,
\end{align}
where the coefficients \(\cZ^{\alpha}(\vect{p})\) parametrise the operator overlaps
\begin{equation}
	\lvac \,\interpoptPEVASi{\mp\vect{p}}{0} | \indSi \,; p\,; s \rangle = \cZ^{\alpha}(\vect{p}) \, \sqrt{\frac{\massSi}{\energySi{\vect{p}}}} \, \Gamma_{\mp\vect{p}} \, \spinorSi{} \, .
\end{equation}

These spinor-projected correlation functions\index{correlation function!three point} have a
nontrivial time dependence, which can be removed by constructing the ratio~\cite{Leinweber:1990dv}\index{ratio}
\begin{align}\label{eqn:formfactors:ratio}
    R^{\indLDm\indLDn}_{\mp}(\vect{p}', \vect{p}\,; \indSj{}, \indSi{}) %\conteqn
    %&\qquad
        &\definedby \, \sqrt{\left|\frac{
        \threecfprojSjSi[\mp]{\vectorcurrentLUm[CI]{}}{\vect{p}'\!}{\vect{p}}{t_2}{t_1}{\projectorLUn{}} \,
        \threecfprojSiSj[\mp]{\vectorcurrentLUm[CI]{}}{\vect{p}}{\vect{p}'\!}{t_2}{t_1}{\projectorLUn{}}}{
        \twocfprojPEVASj{\vect{p}'}{t_2} \, \twocfprojPEVASi{\vect{p}}{t_2}}\right|} \conteqn
    &\quad \times \sign\!\left(
         \threecfprojSjSi[\mp]{\vectorcurrentLUm[CI]{}}{\vect{p}'\!}{\vect{p}}{t_2}{t_1}{\projectorLUn{}}\right) \,,
\end{align}\index{R@\(R_{\mp}(\vect{p}', \vect{p}\,; \alpha\,; r, s)\)|see {ratio}}%
where \(\projectorLUfour{} = (\identity + \gammaLUfour{}) / 2\) and
\(\projectorLCk{} = (\identity + \gammaLUfour{}) (\ii \, \gammaLUfive{} \, \gammaLCk{}) / 2\) form
the basis for the spin projectors we use.

We can then define the reduced ratio\index{reduced ratio}
\begin{align}
    \adjoint{R}^{\indLDm\indLDn}_{\mp}(\vect{p}', \vect{p}\,; \indSj{}, \indSi{}) %\conteqn
    %&\qquad
	&\definedby
        \sqrt{\frac{2 \energySi{\vect{p}}}{\energySi{\vect{p}}+\massSi{}}} \,
        \sqrt{\frac{2 \energySj{\vect{p}'}}{\energySj{\vect{p}'}+\massSi{}}} %\conteqn
    %&\qquad\qquad\quad\times 
    R^{\indLDm\indLDn}_{\mp}(\vect{p}', \vect{p}\,; \indSj{}, \indSi{}) \conteqn
	&= \Tr\!\left[\,\projectorLUn{}\,
	\Gamma_{\mp\vect{p}'}
	\frac{\slashed{p}+\massSi}{\energySi{\vect{p}} + \massSi} \,
	\Gamma^{\mu}_i(p,p') \,
	\frac{\slashed{p}'+\massSj}{\energySj{\vect{p}'} + \massSj}\,
	\Gamma_{+\vect{p}}\,\right]\,.
\label{eq:reducedRatio}
\end{align}

In order to isolate the form factors, we can express this trace (which is itself a linear combination
of the form factors) as a matrix product
\begin{align}
	&\adjoint{R}^{\indLDm\indLDn}_{\mp}(\vect{p}', \vect{p}\,; \indSj{}, \indSi{}) =
	\begin{bmatrix}
		K_{1,\mp}^{\mu\nu}(\vect{p}', \vect{p}\,; \indSj{}, \indSi{} ) & K_{2,\mp}^{\mu\nu}(\vect{p}', \vect{p}\,; \indSj{}, \indSi{} )
	\end{bmatrix}
	\begin{bmatrix}
		F^*_1(Q^2) \\
		F^*_2(Q^2)
	\end{bmatrix} ,
\end{align}
expressed here in terms of the Pauli-Dirac decomposition.
\end{widetext}

The kinematic weights
$K_{i,\mp}^{\mu\nu}(\vect{p}',\vect{p};\indSj, \indSi)$ are determined by substituting the explicit
form for the vertex function in the trace and reducing the product of $\gamma$-matrices into
expressions involving the incoming and outgoing energies and momenta, the eigenstate masses and the
hadron spins via Pauli matrices. The evaluation of these expressions is presented in the
Appendix. By expressing each independent determination of
$\adjoint{R}^{\indLDm\indLDn}_{\mp}(\vect{p}', \vect{p}\,; \indSj{}, \indSi{})$ in this way, we can
combine the results into a single vector equation
\begin{equation}
	\mathbf{R} = \mathbb{K} \, \mathbf{F} \, ,
\end{equation}
where $\mathbf{R}$ is a vector containing $n$ independent ratio determinations, $\mathbb{K}$ is an
$n \times 2$ matrix of kinematic factors and $\mathbf{F}$ is a vector containing the $2$ form
factors. Provided we have at least two linearly independent determinations of
$\overline{R}^{\mu}_{\atobij}(p', p; \Gamma', \Gamma; j, i)$, we can solve the linear system by
taking the pseudo-inverse of $\mathbb{K}$ via the Singular Value Decomposition (SVD) of
$\mathbb{K}$, to give
\begin{equation}
	\mathbf{F} = \mathbb{K}^{+} \, \mathbf{R} \, .
\label{eq:Fsol}
\end{equation}
To express this solution in terms of another choice of form factor decomposition, it is a simple matter of applying the relevant basis transformation matrix relating the two parametrisations. Consequently we can determine the the generalised Sachs form factors and the helicity amplitudes through the following expressions,
\begin{widetext}
\begin{align}
	\mathbf{G} &= 
	\begin{pmatrix}
		G^*_E(Q^2) \\
		G^*_M(Q^2)
	\end{pmatrix} =
	\begin{pmatrix}
		1 & - \frac{Q^2}{(M \pm m)^2} \\	
		1 & 1
	\end{pmatrix}
	\mathbb{K}^{+} \, \mathbf{R}	\, \\
	\mathbf{H} &= 
	\begin{pmatrix}
		{\cal S}_{\frac{1}{2}}(Q^2) \\
		{\cal A}_{\frac{1}{2}}(Q^2)
	\end{pmatrix} =
	b_{\pm}
	\begin{pmatrix}
		\pm \frac{\sqrt{2}(M \pm m)|\q|}{Q^2} & 0 \\	
		0 & 2
	\end{pmatrix}
	\begin{pmatrix}
		1 & - \frac{Q^2}{(M \pm m)^2} \\	
		1 & 1
	\end{pmatrix}
	\mathbb{K}^{+} \, \mathbf{R}	\, .
\label{eq:Hamplitudes}
\end{align}
\end{widetext}

\section{Results\label{sec:results}}

The results presented in this paper are calculated on the PACS-CS \((2+1)\)-flavour full-QCD
ensembles~\cite{PACS-CS:2008bkb}, made available through the ILDG~\cite{Beckett:2009cb}. These
ensembles use a \(32^3 \times 64\) lattice, and employ a renormalisation-group improved Iwasaki
gauge action\index{action:improved} with \(\beta = 1.90\) and non-perturbatively \(O(a)\)-improved
Wilson quarks, with \(\clovercoeff = 1.715\). 

Our current study focuses on the matrix elements of low-lying energy eigenstates at larger quark
masses where the states dominated by single-particle operators are the lowest-lying states in the
finite-volume spectrum.  Our aim is to implement the formalism established in the previous section
and report the first {\it ab initio} calculation of the helicity amplitudes for electromagnetic
transitions from the ground state nucleon to the first two odd-parity excitations.  

The recent HEFT calculation of Ref.~\cite{Abell:2023nex} indicates that the two odd-parity states
observed on the lattice are the lowest states in the spectrum for the heaviest PACS-CS quark mass
ensemble. Moreover, the HEFT analysis indicates scattering state contaminations in the two-point
correlation functions are well below \SI{10}{\percent} for this mass.  Thus of the five PACS-CS
ensembles, we select the ensemble with
\(m_{\pi}=\SI{702}{\mega\electronvolt}\)~\cite{PACS-CS:2008bkb}.  
Given our phenomenological focus is on the quark model, we will consider this particular quark mass
where the assumptions of the non-relativistic constituent quark model are best satisfied.
We set the scale using the Sommer parameter with \(r_0 = \SI{0.4921 \pm
  0.0064}{\femto\meter}\)~\cite{PACS-CS:2008bkb} providing \(a = \SI{0.1022 \pm
  0.0015}{\femto\meter}\) and \( m_\pi^2 = \SI{0.3884 \pm 0.0113}{\giga\electronvolt}^2\) based on
399 configurations.

%% More details of the individual ensembles are presented in Table~\ref{tab:formfactors:ensembles},
%% including the squared pion masses in the Sommer scale. 

For the variational analyses in this paper, we begin with the same
eight-interpolator basis as in Refs.~\cite{Stokes:2018emx,Stokes:2019zdd}, in which we studied
the elastic form factors of the ground-state nucleon and three localised excitations.
This basis is formed from the conventional \spinhalf{} nucleon interpolators
\begin{align}
    \nucleoninterpOne = &\epsilonCaCbCc \, [{\quarkupCa}\transpose \, (C\gammaLUfive) \: \quarkdownCb]\, \quarkupCc\,,\ \text{and}\conteqn
    \nucleoninterpTwo = &\epsilonCaCbCc \, [{\quarkupCa}\transpose \, (C) \: \quarkdownCb]\, \gammaLUfive \quarkupCc\,,
\end{align}
with \num{16}, \num{35}, \num{100}, or \num{200} sweeps of gauge-invariant Gaussian
smearing\index{gauge-invariant Gaussian smearing}~\cite{Gusken:1989qx} with a smearing fraction of
\(\alpha = 0.7\), applied at the quark source and sinks in creating the propagators. For the PEVA
analyses, this basis is expanded to sixteen operators as described in Section~\ref{sec:peva}.  The
gauge links used in the Gaussian smearing are first smoothed by applying four sweeps of
three-dimensional isotropic stout-link smearing~\cite{Morningstar:2003gk} with \(\rho=0.1\).
Using this basis, the electromagnetic transitions between the ground state nucleon and
the first two negative-parity excitations are examined.

To extract the transition form factors, we fix the source at time slice
\(\temporalsites/4=\num{16}\) relative to a fixed boundary condition in time, where $U_t(\vect
x,N_t)=0\,\forall\,{\vect x}$ in the hopping terms of the fermion action.  This prevents
opposite-parity baryon contributions propagating backward in time from wrapping around the lattice
and contaminating our correlations functions.  We work in the middle of the lattice and have
verified that reflections associated with the fixed boundary are negligible for Euclidean times
greater than 16 time slices from the boundary.

Utilising the sequential source technique (SST)~\cite{Bernard:1985ss}, we invert through the
current, fixing the current insertion at time slice \num{21}. We choose time slice \num{21} by
inspecting the two point correlation functions associated with each state and observing that
excited-state contaminations in the eigenstate-projected correlators are suppressed by time slice
\num{21}. This is evaluated by fitting the effective mass\index{effective energy} in this region to
a single-state ansatz verifying that the full correlated \(\chi^2 / \text{dof}\) is
satisfactory. We then extract the transition form factors as outlined in Section~\ref{sec:peva} for
every possible sink time and once again look for a plateau consistent with the single-state ansatz.

When fitting correlators, the \(\chi^2 / \mathrm{dof}\) is calculated with the full covariance
matrix, and the \(\chi^2\) values of all fits are consistent with an appropriate \(\chi^2\)
distribution.  To avoid significant multi-particle contamination, we demand that our two- and
three-point correlators are consistent with a single-state ansatz within the Euclidean time regions
considered. As one proceeds to precision calculations of these transition form factors, it will be
essential to examine the scattering-state contributions to the form factors in detail.

% Q^2

In performing the SST through the electromagnetic current insertion, one needs to commit to specific
momentum insertions, $\vect q$.  In quanta of $2\pi/L$ where $L$ is the length of a side of the
cubic volume, we select three values for $\vect q$ including $(1,0,0)$, $(1,1,0)$ and $(2,0,0)$ for
the three-momentum components.  In the traditional fashion the final momentum of the state is
projected by the Fourier transform of the three-point function, and the initial state momentum is
inferred via momentum conservation.  We consider several initial, $\vect p$, and final ${\vect
p}^\prime$ momenta, with the intention of using boosts to access several momentum transfers
\cite{Stokes:2018emx,Stokes:2019zdd}. For each $\vect q$, we draw from permutations of $\vect p =
(0,0,0)$, $(1,0,0)$, $(1,1,0)$, $(2,0,0)$, $(2,1,0)$, and $(3,0,0)$ with several momentum sign
combinations as we aim to access space-like momentum transfers where $Q^2 = \vect q^{2} - q_0^2$
is positive.  To do so, we ensure the lower-lying nucleon ground state has the largest momentum
between $\vect p$ and $\vect p^\prime$ to minimise the difference $q_0 = p_0^\prime - p_0 = E^\prime
- E$ as the baryons are on-shell.

Of the space-like momentum-transfer combinations considered, four were found to generate good
signal-to-noise correlation functions with excellent plateaus available for fitting.
These momentum combinations are summarised in Table~\ref{tab:momenta}.
Using the masses of the ground-state nucleon, 
      $m_0 = 1.429(09)$ GeV, the first excitation
      $m_1 = 1.996(25)$ GeV, and the second excitation 
      $m_2 = 2.010(39)$ GeV, 
and ensuring the lower-lying nucleon is assigned the larger momentum magnitudes,
the associated $Q^2$ values for the transitions from states $0 \to 1$ and $0 \to 2$ are also
reported in Table~\ref{tab:momenta}.

\begin{table}[tb]
    \caption{The momenta considered in our three-point function analysis
      producing good signal-to-noise in the correlation functions at space-like momentum transfers.
      The initial three-momentum components, $\vect p$, the SST current momentum, $\vect q$, and the
      final momentum, $\vect p^\prime$, are indicated in quanta of $2\pi/L$.  To obtain space-like
      momentum transfers, the lower-lying nucleon is assigned the larger momentum magnitudes of
      $\vect p^\prime$. Using the state masses provided in the text, $Q^2$ values are reported for
      the transitions from states $0 \to 1$ and $0 \to 2$.
%
      %% $m_0 = 1.428817804862191 \pm 0.00908996244344231$, 
      %% $m_1 = 1.996430452122453 \pm 0.02478610496915207$, 
      %% $m_2 = 2.009846418828005 \pm 0.03923682583332354$, 
%
      \label{tab:momenta}
}
\begin{ruledtabular}
\begin{tabular}{ccccc}
\noalign{\smallskip}
\phantom{$-$}$\vect p$ & $\vect q$ & $\vect p^\prime$ & $Q^2_{0 \to 1}/$ GeV & $Q^2_{0 \to 2}/$ GeV \\
\noalign{\smallskip}
\hline
\noalign{\smallskip}
\phantom{$-$}0,0,0 & 1,1,0 & 1,1,0  &0.066(24)  &0.053(38) \\
\phantom{$-$}1,0,0 & 1,1,0 & 2,1,0  &0.150(18)  &0.140(30) \\
          $-$1,0,0 & 2,0,0 & 1,0,0  &0.268(28)  &0.253(44) \\
\phantom{$-$}0,0,0 & 2,0,0 & 2,0,0  &0.431(20)  &0.420(31) \\
%%    0,0,0 & 1,1,0 & 1,1,0  &0.0662346743458571 \pm 0.02432392617681554  &0.0534347067721717 \pm 0.03815069150947662 \\
%%    1,0,0 & 1,1,0 & 2,1,0  &0.1500014598644981 \pm 0.01873225081818004  &0.1400535547212831 \pm 0.02971768154457625 \\
%% $-$1,0,0 & 2,0,0 & 1,0,0  &0.2681486378289932 \pm 0.02827218027089553  &0.2533732235088975 \pm 0.04399347093807013 \\
%%    0,0,0 & 2,0,0 & 2,0,0  &0.4313361033559904 \pm 0.01950048375213905  &0.4209896089852103 \pm 0.03090812268655191 \\
\end{tabular}
\end{ruledtabular}
\end{table}

It's interesting to note that $\vect q = (1,0,0)$ is insufficient to generate space-like momentum
transfers.  The mass splitting between the states is significant and larger momentum differences
are required to bring the state energies closer together.

To this point our focus has been on addressing the issue of parity mixing via the PEVA technique.
This is particularly important for odd-parity excitations where the even-parity ground state can
enter and generate systematic errors that grow at a large exponential rate.

However, for hadrons in flight, momentum and angular momentum are no longer simultaneous
eigenvalues and the issue of angular momentum mixing enters.  Indeed, experiment considers a
particle in its centre of momentum (CoM) frame to describe its angular momentum.  As discussed
above, access to space-like momentum transfers in the electromagnetic transition requires at least
one of the baryons to be in flight.

This problem of angular momentum mixing is exacerbated by the lattice subduction of the continuous
spatial symmetries of the rotational group to the discrete symmetries of the cubic group.

At zero momentum, the relevant symmetry group is ${}^2O$, the double cover of the cubic group
\cite{Johnson:1982yq}.  Here the continuum $J^P = 1/2^\pm$ subduce to the $G_1^{g/u}$ irreducible
representations (irreps) where the next value of $J$ that appears is $7/2$.  In light of the large
gap in angular momenta, the use of continuum symmetries is considered effective for isolating
$J=1/2^\pm$ states at rest \cite{Meinel:2020owd}.

However, at nonzero momenta, the relevant symmetry groups are the Little Groups of ${}^2O$
\cite{Gockeler:2012yj,Morningstar:2013bda}, and the gap in angular momentum mixing is lost. For
example, an irrep can contain both $J=1/2$ and $J=3/2$.  Here a finite-volume state associated with
the $J^P=3/2^-$ $N^*(1520)$ is of particular concern as it is nearly degenerate with the $J^P=1/2^-$
states of interest.

Thus, it is advantageous to excite the odd-parity spin-1/2 state at rest and have the low-lying
even-parity state carry finite momentum.  In this case Euclidean time evolution can be exploited to
suppress the spin-3/2 contamination from the even-parity ground-state nucleon.  Fortunately, this
preference is in accord with accessing spatial momentum transfers.  Both the smallest and largest
momentum transfers considered are obtained with the odd-parity excitation at rest.

The two intermediate momentum transfers considered in Table \ref{tab:momenta} have both initial-
and final-state baryons in flight.  However, the minimum non-trivial momentum is applied to the
odd-parity excitation.  While our lattice setup is designed to minimise the $J=3/2$ contamination,
it will be important to address this in a next generation calculation.

As one might expect, when the excited state takes zero momentum, one obtains good signal to noise
in the three-point correlation function. Another favourable momentum combination is when the
momenta of the incoming and outgoing states differ by a sign.  It's well known this type of Breit
frame kinematics is favourable for form factors, and we now see it to be favourable in the case of
transition form factors.  The fourth favourable momentum combination in the second row of
Table~\ref{tab:momenta} is perhaps somewhat of a surprise.  However, the three-momentum of the
current insertion is the smallest available to generate space-like momentum transfer, and the
magnitude of $Q^2$ is the second smallest of those considered.

In the next two subsections we examine the Pauli-Dirac form factors $F_1^*(Q^2)$ and $F_2^*(Q^2)$
for electromagnetic transitions from the ground-state nucleon to the first and second negative
parity excitations.  These form factors are defined in the covariant vertex function of
Eq.~(\ref{eq:abnormtransF1F2}) and used in Eq.~(\ref{eq:pchanging}).  Subsequently, they will be
related to the helicity amplitudes via the kinematic factors of
Eq.~(\ref{eq:Hamplitudes}). However, the fits to our correlation functions are performed for
$F_1^*(Q^2)$ and $F_2^*(Q^2)$.

The form factors are obtained in the solution of Eq.~(\ref{eq:Fsol}) where the vector $\mathbf{R}$
is constructed from the nontrivial components of the $4 \times 4 \times 2$ elements of
$\adjoint{R}^{\indLDm\indLDn}_{\mp}(\vect{p}', \vect{p}\,; \indSj{}, \indSi{})$ defined in
Eq.~(\ref{eq:reducedRatio}).  We report contributions from the doubly-represented $u$-sector of the
proton, $u_p$, and the singly-represented $d$-sector of the proton, $d_p$, individually.  Each
sector is reported for single quarks of unit charge such that the physical states are obtained
through simple quark counting and electric charge assignments.  Schematically, the physical proton
and neutron transition amplitudes are obtained via
\begin{subequations}
\label{eq:phys} 
\begin{align}
p &= \phantom{-}\frac{2}{3}\, 2 \, u_p - \frac{1}{3} \, d_p \, , \\
n &=           -\frac{1}{3}\, 2 \, u_p + \frac{2}{3} \, d_p \, .
\end{align}
\end{subequations}

\subsection{Transitions to the first negative parity excitation\label{sec:results:1stneg}}

\begin{figure}[t!]
    {\centering
	\resizebox{1.0\columnwidth}{!}{\includegraphics{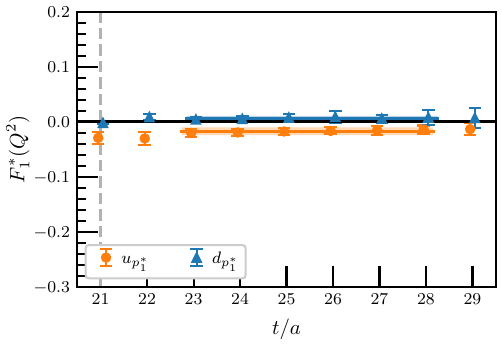}}}
    \vspace{-16pt}
    \caption{\label{fig:plateau.state0to1.f1.p000.pp110} Quark-sector contributions to the
      three-point correlation function for the covariant vertex function $F_1^*(Q^2)$ at $Q^2 =
      0.066(24)$ GeV${}^2$ for the odd-parity transition between the ground state proton and the
      first negative-parity excitation. Sector contributions are normalised to represent single
      quarks of unit charge. }
\end{figure}

\begin{figure}[t!]
    {\centering
        \resizebox{1.0\columnwidth}{!}{\includegraphics{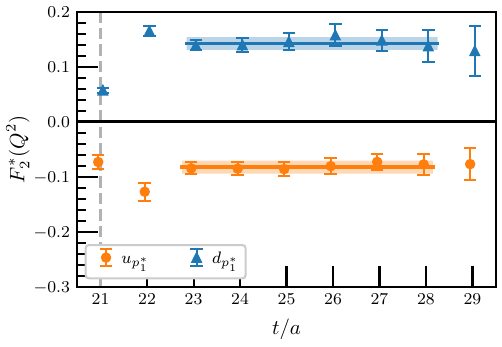}}}
    \vspace{-16pt}
    \caption{\label{fig:plateau.state0to1.f2.p000.pp110} Quark-sector contributions to the
      three-point correlation function for the covariant vertex function $F_2^*(Q^2)$ at $Q^2 =
      0.066(24)$ GeV${}^2$ for the odd-parity transition between the ground state proton and the
      first negative-parity excitation. Sector contributions are normalised to represent single
      quarks of unit charge. }
\end{figure}

\begin{figure}[t!]
    {\centering
        \resizebox{1.0\columnwidth}{!}{\includegraphics{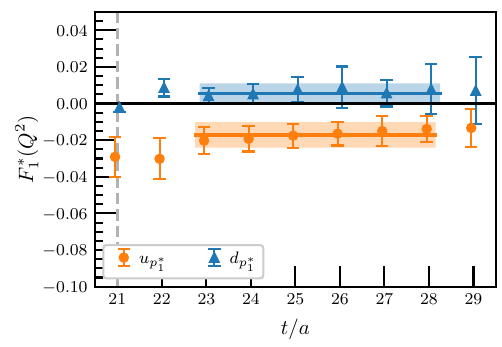}}}
    \vspace{-16pt}
    \caption{\label{fig:plateau.state0to1.f1.p000.pp110.zoom} Zoomed-in presentation of the lattice
      QCD results presented in Fig.~\ref{fig:plateau.state0to1.f1.p000.pp110}
      %% Quark-sector contributions to the three-point correlation function for the covariant
      %% vertex function $F_1^*(Q^2)$ at $Q^2 = 0.066(24)$ GeV${}^2$ 
      for the transition between the ground state proton and the first negative-parity
      excitation. The onset of plateau behaviour is observed at two time steps following the
      current insertion at $t=21$ and statistical uncertainties remain stable throughout the fit
      regime.}
\end{figure}

Results for $F_1^*(Q^2)$ and $F_2^*(Q^2)$ in the odd parity electromagnetic transition for the
ground-state proton to the first negative parity excitation are presented in
Figs.~\ref{fig:plateau.state0to1.f1.p000.pp110} and \ref{fig:plateau.state0to1.f2.p000.pp110}
respectively.  Here $Q^2$ takes the value 0.066(24) GeV${}^2$, the lowest space-like momentum
transfer available in this analysis.  Values for $F_1^*$ are small and a zoomed-in presentation of
the fit is provided in Fig.~\ref{fig:plateau.state0to1.f1.p000.pp110.zoom}.

The use of correlation matrix eigenvectors to aid in isolating the energy eigenstates leads to
a rapid onset of plateau behaviour following the current insertion at $t=21$.  By two-time
slices later, a persistent plateau is identified in the full covariance matrix fit.

It's interesting that for both form factors, the doubly-represented $u$ sector of the proton
contributes with a negative sign, while the singly-represented sector contributes with an opposite
positive sign.  In the dominant $F_2^*$ form factor, the singly-represented $d$ sector of the
proton makes the largest contribution.  Nevertheless, upon including quark-counting and charge
factors, the doubly-represented $u$-quark sector of the proton will dominate, whereas the singly
represented $u$-quark sector in the neutron will dominate.

\begin{figure}[t!]
    {\centering
        \resizebox{1.01\columnwidth}{!}{\includegraphics{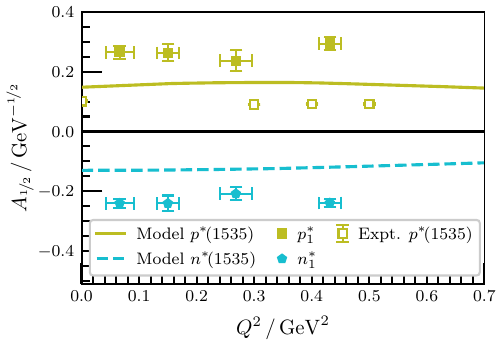}}}
    \vspace{-16pt}
    \caption{\label{fig:1stneg:A:k1}Transverse helicity amplitudes for transitions from the ground
      state nucleon to the first negative-parity excitation at \(m_{\pi} =
      \SI{702}{\mega\electronvolt}\) (filled symbols) are presented in the context of results from
      a constituent quark model incorporating relativistic effects in a light-front framework
      \cite{Capstick:1994ne} (lines) and experiment (open symbols).  }
\end{figure}

\begin{figure}[t!]
    {\centering
        \resizebox{1.01\columnwidth}{!}{\includegraphics{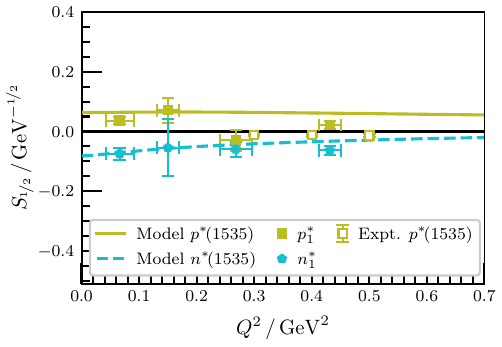}}}
    \vspace{-16pt}
    \caption{\label{fig:1stneg:S:k1}Longitudinal helicity amplitudes for transitions from the ground
      state nucleon to the first negative-parity excitation at \(m_{\pi} =
      \SI{702}{\mega\electronvolt}\) (filled symbols) are presented in the context of results from
      a relativised constituent quark model 
      \cite{Capstick:1994ne} (lines) and experiment (open symbols).  }
\end{figure}

\begin{figure}[t!]
    {\centering
        \resizebox{1.01\columnwidth}{!}{\includegraphics{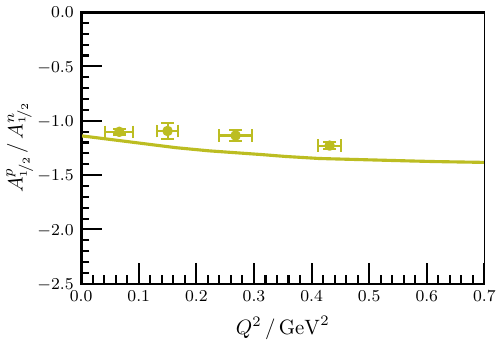}}}
    \vspace{-16pt}
    \caption{\label{fig:1stneg:A:ratio:k1}Ratio of proton to neutron transverse helicity amplitudes
      for transitions from the ground state to the first negative-parity excitation at \(m_{\pi} =
      \SI{702}{\mega\electronvolt}\) (filled symbols) are compared with the relativised constituent
      quark model \cite{Capstick:1994ne}.  }
\end{figure}

With $F_1^*(Q^2)$ and $F_2^*(Q^2)$ determined in the correlation function ratio fits, we proceed to
calculate the helicity amplitudes for the ground to first odd-parity excitation.  Quark sectors are
combined to create physical proton and neutron states via Eqs.~(\ref{eq:phys}) and the helicity
amplitudes are obtained via Eq.~(\ref{eq:Hamplitudes}).  Figures~\ref{fig:1stneg:A:k1} and
\ref{fig:1stneg:S:k1} present the transverse and longitudinal helicity amplitudes respectively, as
a function of the momentum transfer, $Q^2$.  Here our lattice QCD results at $m_{\pi} =
\SI{702}{\mega\electronvolt}$ are presented in the context of constituent quark model predictions
for the $N^*(1535)$ and experimental measurements.  The quark model \cite{Capstick:1994ne}
incorporates relativistic effects in a light-front framework.

Our first observation for the transverse helicity amplitude is that the $Q^2$ dependence is very
mild, a result reflected in both the constituent quark model and experiment.  And while the sign of
the proton and neutron helicity amplitudes are in accord with the model and available
experimental results, the magnitude of our lattice results exceed both the model and experiment.
It will be interesting to learn the relevant dynamics as one reduces the quark masses of the
lattice QCD simulation toward the physical point. Assuming the lattice results approach the
experimental measurements, it will be interesting to understand the relative roles of a simple
quark mass dependence for each state contributing to the resonance amplitude and the
resonance-related role of several finite-volume states contributing to the amplitude.

The longitudinal helicity amplitudes presented in Fig.~\ref{fig:1stneg:S:k1} are relatively small.
Here good agreement with the model results is observed for the excited proton at small $Q^2$ but
the lattice QCD results do not show the $Q^2$ invariance of the model, instead dropping towards zero
and even changing sign in accord with the experimental measurements.  On the other hand, the
lattice QCD results agree very well with the model results for the transition of the neutron to the
$n^*(1535)$. Together, these results suggest that the quark mass dependence of this observable is
likely to be mild.

It's interesting to examine the extent to which the lattice QCD and model predictions of the
helicity amplitudes agree after the ratio of proton to neutron amplitudes are taken.  In the ratio,
quark mass effects have an opportunity to compensate, leaving a prediction more robust to quark
mass variation.  Figure \ref{fig:1stneg:A:ratio:k1} presents this ratio for the transverse helicity
amplitude as a function of $Q^2$.  Again the trends match and the magnitudes are now quite similar.

\subsection{Transitions to the second negative parity excitation\label{sec:results:2ndneg}}

We now turn our attention to the electromagnetic transitions to the second low-lying
negative-parity nucleon excitation observed on the lattice at $m_{\pi} =
\SI{702}{\mega\electronvolt}$ and compare our results with quark model calculations of the
$N^*(1650)$ and associated experimental measurements.  It is here that the correlation matrix based
eigenstate projection gains significant importance as we examine electromagnetic transitions to
the second eigenstate in the odd-parity spectrum.

\begin{figure}[t!]
    {\centering
        \resizebox{1.0\columnwidth}{!}{\includegraphics{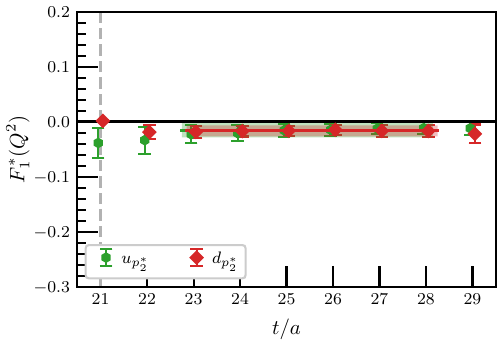}}}
    \vspace{-16pt}
    \caption{\label{fig:plateau.state0to2.f1.p000.pp110} Quark-sector contributions to the
      three-point correlation function for the covariant vertex function $F_1^*(Q^2)$ at $Q^2 =
      0.053(38)$ GeV${}^2$ for the odd-parity transition between the ground state proton and the
      second negative-parity excitation. Sector contributions are normalised to represent single
      quarks of unit charge. }
\end{figure}

\begin{figure}[t!]
    {\centering
        \resizebox{1.0\columnwidth}{!}{\includegraphics{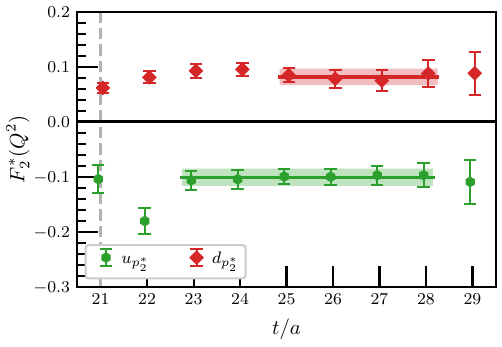}}}
    \vspace{-16pt}
    \caption{\label{fig:plateau.state0to2.f2.p000.pp110} Quark-sector contributions to the
      three-point correlation function for the covariant vertex function $F_2^*(Q^2)$ at $Q^2 =
      0.053(38)$ GeV${}^2$ for the odd-parity transition between the ground state proton and the
      second negative-parity excitation. Sector contributions are normalised to represent single
      quarks of unit charge. }
\end{figure}

\begin{figure}[t!]
    {\centering
        \resizebox{1.0\columnwidth}{!}{\includegraphics{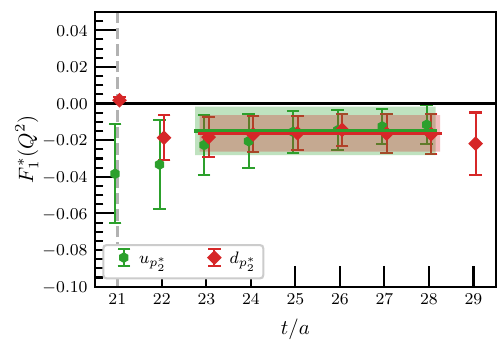}}}
    \vspace{-16pt}
    \caption{\label{fig:plateau.state0to2.f1.p000.pp110.zoom} Zoomed-in presentation of the lattice
      QCD results presented in Fig.~\ref{fig:plateau.state0to2.f1.p000.pp110} 
      %% Quark-sector contributions to the three-point correlation function for the covariant
      %% vertex function $F_1^*(Q^2)$ at $Q^2 = 0.053(38)$ GeV${}^2$
      for the transition between the ground state proton and the second negative-parity
      excitation. Statistical uncertainties are larger for this transition relative to the
      transition to the first negative-parity excitation illustrated in
      Fig.~\ref{fig:plateau.state0to1.f1.p000.pp110.zoom}.}
\end{figure}

Results for $F_1^*(Q^2)$ and $F_2^*(Q^2)$ in the electromagnetic transition from the ground-state
proton to the second negative parity excitation are presented in
Figs.~\ref{fig:plateau.state0to2.f1.p000.pp110} and \ref{fig:plateau.state0to2.f2.p000.pp110}
respectively.  Here the minimum space-like $Q^2$ takes a slightly smaller value of 0.053(38)
GeV${}^2$ due to the change in the mass of the excitation.  Once again, values for $F_1^*$ are
small and a zoomed-in presentation of the fit is provided in
Fig.~\ref{fig:plateau.state0to2.f1.p000.pp110.zoom}.  Here we observe slightly larger statistical
uncertainties for the transition to the second odd-parity excitation.

The use of correlation-matrix eigenvector projection for state isolation leads to a rapid onset of
plateau behaviour for three of the four correlators presented.  Some fluctuation in the proton's
singly-represented $d$-quark sector contribution to $F_2^*(Q^2)$ prevents an early fit due to
uncomfortably large values for the full covariance matrix $\chi^2/$dof.

This time there is a clear sign change between the contributions to $F_1^*(Q^2)$ and $F_2^*(Q^2)$
for the singly-represented $d$-quark sector in the proton.  However, the doubly-represented $u$
sector of the proton continues to uniformly contribute with a negative sign. In the dominant
$F_2^*$ form factor, both quark sectors make a contribution that is similar in magnitude.  Upon
including quark-counting and charge factors, the $u$-quark sector of the proton will dominate,
whereas the sectors are more balanced for the neutron.

\begin{figure}[t!]
    {\centering
        \resizebox{1.01\columnwidth}{!}{\includegraphics{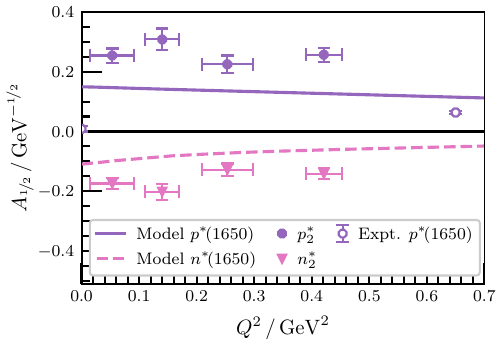}}}
    \vspace{-16pt}
    \caption{\label{fig:2ndneg:A:k1}Transverse helicity amplitudes for transitions from the ground
      state nucleon to the second negative-parity excitation at \(m_{\pi} =
      \SI{702}{\mega\electronvolt}\) (filled symbols) are presented in the context of results from
      a constituent quark model incorporating relativistic effects in a light-front framework
      \cite{Capstick:1994ne} (lines) and experiment (open symbols).  }
\end{figure}

\begin{figure}[t!]
    {\centering
        \resizebox{1.01\columnwidth}{!}{\includegraphics{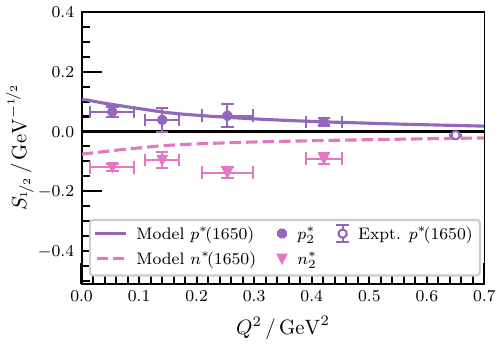}}}
    \vspace{-16pt}
    \caption{\label{fig:2ndneg:S:k1}Longitudinal helicity amplitudes for transitions from the
      ground state nucleon to the second negative-parity excitation at \(m_{\pi} =
      \SI{702}{\mega\electronvolt}\) (filled symbols) are presented in the context of results from
      a relativised constituent quark model \cite{Capstick:1994ne} (lines) and experiment (open
      symbols).  }
\end{figure}

\begin{figure}[t!]
    {\centering
        \resizebox{1.01\columnwidth}{!}{\includegraphics{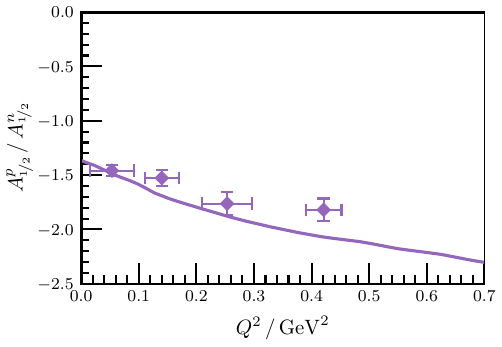}}}
    \vspace{-16pt}
    \caption{\label{fig:2ndneg:A:ratio:k1}Ratio of proton to neutron transverse helicity amplitudes
      for transitions from the ground state to the second negative-parity excitation at \(m_{\pi} =
      \SI{702}{\mega\electronvolt}\) (filled symbols) are compared with the relativised constituent
      quark model \cite{Capstick:1994ne}.  }
\end{figure}

With $F_1^*(Q^2)$ and $F_2^*(Q^2)$ determined, we proceed to the helicity amplitudes for the ground
to second odd-parity excitation.  Figures~\ref{fig:2ndneg:A:k1} and \ref{fig:2ndneg:S:k1} present
the transverse and longitudinal helicity amplitudes respectively.  Our lattice QCD results are
presented in the context of the relativised constituent quark model \cite{Capstick:1994ne} and
experimental measurements where available.

A similar pattern of results for the transverse helicity amplitude is observed.  Again, the $Q^2$
dependence is very mild in both the lattice and constituent quark model results.  The signs of the
proton and neutron helicity amplitudes from lattice QCD are in accord with the model and
experiment.  Again, the magnitude of our lattice results exceed both the model and experiment,
suggesting an important quark mass dependence.

The longitudinal helicity amplitudes presented in Fig.~\ref{fig:2ndneg:S:k1} are relatively small.
Reasonable agreement with the model results are observed.  However, the sign of the experimental
result at larger $Q^2$ differs.  A high-statistics analysis would be required to access this $Q^2$,
but such an analysis would be able to discern if the sign change is $Q^2$ related or quark-mass
related.

On a final note, we examine the extent to which the lattice QCD and model predictions of the
transverse helicity amplitudes agree after the ratio of proton to neutron amplitudes are taken to
make a comparison more robust to quark mass variation.  Figure \ref{fig:2ndneg:A:ratio:k1}
illustrates matching $Q^2$ trends with comparable magnitudes in the ratio.

\section{Conclusion\label{sec:summary}}

A formalism for the determination of spin-1/2 electromagnetic transition form factors in lattice
QCD has been established.  Both normal parity-conserving transitions and abnormal parity-changing
transitions have been included in our considerations.  After working with the standard covariant
vertex functions, we introduce a new set of covariant vertex functions which more closely resemble
the Pauli-Dirac elastic form factors.

Using the parity-expanded variational analysis (PEVA) technique
\cite{Stokes:2013fgw,Stokes:2018emx,Stokes:2019zdd}, the new formalism is implemented in the first
lattice QCD calculation of the helicity amplitudes for transitions from the ground state nucleon to
both the first and the second odd-parity excitations observed in lattice QCD at \(m_{\pi} =
\SI{702}{\mega\electronvolt}\).  The selection of a large pion mass avoids significant
complications associated with the mixing with two-particle states to become scattering states.

Statistical uncertainties in the transition form factors are found to be similar for both states,
with only slightly larger fluctuations in the Pauli-Dirac form factor $F_1^*(Q^2)$ and slightly
larger statistical uncertainties in $F_2^*(Q^2)$ for the second odd-parity excitation of the
nucleon.

Our lattice QCD results are compared comprehensively with a relativised constituent quark model
using a light-front formalism \cite{Capstick:1994ne}.  In every case, the lattice QCD results for
the transverse and longitudinal helicity amplitudes agree with the quark model in sign.  This
includes both proton and neutron amplitudes and electromagnetic transitions to both the first and
the second excitation.  Moreover, when a ratio of proton to neutron results is considered to
suppress quark-mass related effects, the results compare favourably.  

We also considered issues of angular momentum mixing for in-flight baryons associated with the
subduction of continuous spatial symmetries to the Little Groups of the double cover of the cubic
group, ${}^2O$. While the smallest and largest momentum transfers considered draw on Euclidean time
evolution to suppress $J=3/2$ contaminations from the even-parity ground-state nucleon, the two
intermediate momentum transfers considered are susceptible to $J=3/2$ state contamination. Although
our lattice setup is designed to minimise these contaminations, it will be important to address
this in a next generation calculation.

This issue of angular-momentum mixing can be addressed using the techniques presented here for
parity.  This time, one expands the lattice correlation matrix for two-point functions by including
$J=3/2$ interpolating fields.  These interpolating fields can then enter in the process of
isolating $J=1/2$ energy eigenstates by removing $J=3/2$ contaminations from the $J=1/2$ CoM
states.  With the GEVP eigenvectors determined, one can then apply them to the three-point
correlations functions, as we have done herein.

Combined with our earlier results favourably comparing lattice QCD and the quark model for the
electromagnetic form factors of these odd-parity excitations, the results presented here provide
further credence that the two low-lying odd-parity excitations of the nucleon observed
in lattice QCD at $m_\pi \simeq 700$ MeV are associated with quark-model like
basis states which are dressed to form resonances.  With the presence of
  single-particle basis states established, Hamiltonian effective field theory shows that these
  degrees of freedom make an important contribution to experimental scattering amplitudes
  \cite{Abell:2023nex} and photoproduction of the low-lying odd-parity resonances
  \cite{Zhuge:2024iuw}.

The magnitude of the lattice QCD results for the transverse helicity amplitudes are larger than
that predicted by the quark model as it was tuned to describe experiment.  Indeed, the experimental
measurements are even smaller than that predicted by the quark model.  Together, these hint at a
quark mass effect that is significant for the transverse helicity amplitude.  It will be
interesting to learn the relevant dynamics as one reduces the quark masses of the lattice QCD
simulation toward the physical point. While the predominant effect may be a simple quark-mass
dependence for each state contributing to the resonance amplitude, effects from combining the
contributions from several finite-volume states to form the resonance are a novel possibility.

Thus working towards the physical quark-mass requires the introduction of multi-particle
interpolating fields that will mix with the single-particle interpolators considered herein
\cite{Abell:2023nex}.  In the three-point function calculation, one will need to perform current
insertions on all five of the quark propagators, further increasing the demands of the calculation.
As the bare quark-model-like basis state mixes with the nearby two-particle states its contribution
will be spread over several finite-volume states.  It will be fascinating to learn the transition
amplitudes to each of these finite-volume states and how they differ.  Finally, it will be
important to probe larger space-like $Q^2$ values to make direct confrontations with experimental
measurements.

\appendix*
\begin{widetext}
\section{Kinematic factors}
As discussed in Section~\ref{sec:threepoint}, the reduced ratio
\(\adjoint{R}^{\indLDm\indLDn}_{\mp}(\vect{p}', \vect{p}\,; \indSj{}, \indSi{})\) can
be expressed as a matrix product
\begin{align}
	&\adjoint{R}^{\indLDm\indLDn}_{\mp}(\vect{p}', \vect{p}\,; \indSj{}, \indSi{}) =
	\begin{bmatrix}
		K_{1,\mp}^{\mu\nu}(\vect{p}', \vect{p}\,; \indSj{}, \indSi{} ) & K_{2,\mp}^{\mu\nu}(\vect{p}', \vect{p}\,; \indSj{}, \indSi{} )
	\end{bmatrix}
	\begin{bmatrix}
		F^*_1(Q^2) \\
		F^*_2(Q^2)
	\end{bmatrix} ,
\end{align}
of some kinematic weights \(K_{i,\pm}^{\mu\nu}(\vect{p}',\vect{p};\indSj, \indSi)\) and the
Pauli-Dirac-like form factors, \(F^*_1(Q^2)\) and \(F^*_2(Q^2)\). These kinematic factors can
be obtained by evaluating traces of the form
\begin{equation}
    F'_{\mp}\left(\projector{}, \current\right) \definedby 8 E E'
        \Tr\!\left( \projector{} \, \projmpp \, \frac{-\ii \gamma \cdot p' + \identity M'}{2 E'} \,
        \current \, \frac{-\ii \gamma \cdot p + \identity M}{2 E} \, \projm \right)\,.
\end{equation}

Specifically, in the Pauli representation, for normal, parity-conserving transitions,
\begin{align}
	K_{1,\mp,n}^{\mu\nu}(\vect{p}',\vect{p};\indSj, \indSi) &=
		\left(\delta^{\mu\rho}{} - \frac{\indexLDm{q}q^{\rho}}{q^2}\right) F'_{\mp}(\projectorLUn{}, \gamma^\rho) \,\text{ and}\\
	K_{2,\mp,n}^{\mu\nu}(\vect{p}',\vect{p};\indSj, \indSi) &= 
		- \frac{q^\rho}{\massSj + \massSi{}} \, F'_{\mp}(\projectorLUn{}, \sigma^{\mu\rho}) \,,
\end{align}
and for abnormal, parity-changing transitions
\begin{align}
	K_{1,\mp,a}^{\mu\nu}(\vect{p}',\vect{p};\indSj, \indSi) &=
		\left(\delta^{\mu\rho}{} - \frac{\indexLDm{q}q^{\rho}}{q^2}\right) F'_{\mp}(\projectorLUn{}, \gamma^\rho \gamma^5) \,\text{ and}\\
	K_{2,\mp,a}^{\mu\nu}(\vect{p}',\vect{p};\indSj, \indSi) &=
		- \frac{q^\rho}{\massSj - \massSi{}} \, F'_{\mp}(\projectorLUn{}, \sigma^{\mu\rho} \gamma^5) \,.
\end{align}

These traces were evaluated in the Pauli representation in Ref.~\cite{Stokes:2018mki}, and are
reproduced below for completeness:
\begin{align*}
    F'_{\mp}\left({\projectorLUfour{}}, {\identity{{}}}\right)
        &= \left(\left(E + m\right) \left(E' + m'\right) \mp |\vect{p}| |\vect{p}'|\right) \left(1 \pm \hat{p}^{i} \hat{p}^{\prime\,i}\right) \neweqn
    F'_{\mp}\left({\projectorLUfour{}}, {\gamma^{i}}\right)
        &= -\left(|\vect{p}| \left(E' + m'\right) \pm |\vect{p}'| \left(E + m\right)\right) \left(\pm \epsilon^{ijk} \hat{p}^{j} \hat{p}^{\prime\,k} + {\ii} \left(\hat{p}^{i} \pm \hat{p}^{\prime\,i}\right)\right) \neweqn
    F'_{\mp}\left({\projectorLUfour{}}, {\gamma^{4}}\right)
        &= \left(\left(E + m\right) \left(E' + m'\right) \pm |\vect{p}| |\vect{p}'|\right) \left(1 \pm \hat{p}^{i} \hat{p}^{\prime\,i}\right) \neweqn
    F'_{\mp}\left({\projectorLUfour{}}, {\gamma^5}\right)
        &= \left(|\vect{p}| \left(E' + m'\right) \mp |\vect{p}'| \left(E + m\right)\right) \left(1 \pm \hat{p}^{i} \hat{p}^{\prime\,i}\right) \neweqn
    F'_{\mp}\left({\projectorLUfour{}}, {\gamma^{i} \gamma^5}\right)
        &= -\left(\left(E + m\right) \left(E' + m'\right) \pm |\vect{p}| |\vect{p}'|\right) \left(\pm \epsilon^{ijk} \hat{p}^{j} \hat{p}^{\prime\,k} + {\ii} \left(\hat{p}^{i} \pm \hat{p}^{\prime\,i}\right)\right) \neweqn
    F'_{\mp}\left({\projectorLUfour{}}, {\gamma^{4} \gamma^5}\right)
        &= \left(|\vect{p}| \left(E' + m'\right) \pm |\vect{p}'| \left(E + m\right)\right) \left(1 \pm \hat{p}^{i} \hat{p}^{\prime\,i}\right) \neweqn
    F'_{\mp}\left({\projectorLUfour{}}, {\sigma^{4i}}\right)
        &= {\ii} \left(|\vect{p}| \left(E' + m'\right) \mp |\vect{p}'| \left(E + m\right)\right) \left(\pm \epsilon^{ijk} \hat{p}^{j} \hat{p}^{\prime\,k} + {\ii} \left(\hat{p}^{i} \pm \hat{p}^{\prime\,i}\right)\right) \neweqn
    F'_{\mp}\left({\projectorLUfour{}}, {\sigma^{ij}}\right)
        &= -\left(\left(E + m\right) \left(E' + m'\right) \mp |\vect{p}| |\vect{p}'|\right) \left(\epsilon^{ijk} \left(\hat{p}^{k} \pm \hat{p}^{\prime\,k}\right) \mp {\ii} \left(\hat{p}^{i} \hat{p}^{\prime\,j} - \hat{p}^{j} \hat{p}^{\prime\,i}\right)\right) \neweqn
    F'_{\mp}\left({\projectorLCk{}}, {\identity{{}}}\right)
        &= {-\ii} \left(\left(E + m\right) \left(E' + m'\right) \mp |\vect{p}| |\vect{p}'|\right) \left(\pm \epsilon^{ijk} \hat{p}^{j} \hat{p}^{\prime\,k} - {\ii} \left(\hat{p}^{i} \pm \hat{p}^{\prime\,i}\right)\right) \neweqn
    F'_{\mp}\left({\projectorLCk{}}, {\gamma^{j}}\right)
        &= {\ii} \left(|\vect{p}| \left(E' + m'\right) \pm |\vect{p}'| \left(E + m\right)\right) \conteqn
        &\qquad \times \left(\pm \hat{p}^{i} \hat{p}^{\prime\,j} \pm \hat{p}^{j} \hat{p}^{\prime\,i} - {\ii} \epsilon^{ijk} \left(\hat{p}^{k} \mp \hat{p}^{\prime\,k}\right) + \delta^{ij} \left(1 \mp \hat{p}^{k} \hat{p}^{\prime\,k}\right)\right) \neweqn
    F'_{\mp}\left({\projectorLCk{}}, {\gamma^{4}}\right)
        &= {-\ii} \left(\left(E + m\right) \left(E' + m'\right) \pm |\vect{p}| |\vect{p}'|\right) \left(\pm \epsilon^{ijk} \hat{p}^{j} \hat{p}^{\prime\,k} - {\ii} \left(\hat{p}^{i} \pm \hat{p}^{\prime\,i}\right)\right) \neweqn
    F'_{\mp}\left({\projectorLCk{}}, {\gamma^5}\right)
        &= {-\ii} \left(|\vect{p}| \left(E' + m'\right) \mp |\vect{p}'| \left(E + m\right)\right) \left(\pm \epsilon^{ijk} \hat{p}^{j} \hat{p}^{\prime\,k} - {\ii} \left(\hat{p}^{i} \pm \hat{p}^{\prime\,i}\right)\right) \neweqn
    F'_{\mp}\left({\projectorLCk{}}, {\gamma^{j} \gamma^5}\right)
        &= {\ii} \left(\left(E + m\right) \left(E' + m'\right) \pm |\vect{p}| |\vect{p}'|\right) \conteqn
        &\qquad \times \left(\pm \hat{p}^{i} \hat{p}^{\prime\,j} \pm \hat{p}^{j} \hat{p}^{\prime\,i} - {\ii} \epsilon^{ijk} \left(\hat{p}^{k} \mp \hat{p}^{\prime\,k}\right) + \delta^{ij} \left(1 \mp \hat{p}^{k} \hat{p}^{\prime\,k}\right)\right) \neweqn
    F'_{\mp}\left({\projectorLCk{}}, {\gamma^{4} \gamma^5}\right)
        &= {-\ii} \left(|\vect{p}| \left(E' + m'\right) \pm |\vect{p}'| \left(E + m\right)\right) \left(\pm \epsilon^{ijk} \hat{p}^{j} \hat{p}^{\prime\,k} - {\ii} \left(\hat{p}^{i} \pm \hat{p}^{\prime\,i}\right)\right) \neweqn
    F'_{\mp}\left({\projectorLCk{}}, {\sigma^{4j}}\right)
        &= \left(|\vect{p}| \left(E' + m'\right) \mp |\vect{p}'| \left(E + m\right)\right) \conteqn
        &\qquad \times \left(\pm \hat{p}^{i} \hat{p}^{\prime\,j} \pm \hat{p}^{j} \hat{p}^{\prime\,i} - {\ii} \epsilon^{ijk} \left(\hat{p}^{k} \mp \hat{p}^{\prime\,k}\right) + \delta^{ij} \left(1 \mp \hat{p}^{k} \hat{p}^{\prime\,k}\right)\right) \neweqn
    F'_{\mp}\left({\projectorLCk{}}, {\sigma^{jk}}\right)
        &= \left(\left(E + m\right) \left(E' + m'\right) \mp |\vect{p}| |\vect{p}'|\right) \epsilon^{jkl} \conteqn
        &\qquad \times \left(\pm \hat{p}^{i} \hat{p}^{\prime\,l} \pm \hat{p}^{l} \hat{p}^{\prime\,i} - {\ii} \epsilon^{ilm} \left(\hat{p}^{m} \mp \hat{p}^{\prime\,m}\right) + \delta^{il} \left(1 \mp \hat{p}^{m} \hat{p}^{\prime\,m}\right)\right)
\end{align*}

\newpage

\end{widetext}

\begin{acknowledgments}
We thank the PACS-CS Collaboration for making their configurations available via the International
Lattice Data Grid (ILDG).  
This research was undertaken with the assistance of resources from the Phoenix HPC service at the
University of Adelaide, the National Computational Infrastructure (NCI), which is supported by the
Australian Government, and by resources provided by the Pawsey Supercomputing Centre with funding
from the Australian Government and the Government of Western Australia. These resources were
provided through the National Computational Merit Allocation Scheme and the University of Adelaide
partner share.
FS is supported by a Ramsay Fellowship from the University of Adelaide.
WK was supported by the Pawsey Supercomputing Centre through the Pawsey Centre for Extreme Scale
Readiness (PaCER) program.
This research is supported by the Australian Research Council through grants no.\ DP140103067,
DP150103164, LE160100051, LE190100021, DP190102215, and DP210103706.
\end{acknowledgments}

\bibliography{refs}

\end{document}